\documentclass[aoas,MSNbibl,nameyear,seceqn,dvips]{arximspdf}
\usepackage{mathbh,multirow,dcolumn}
\usepackage{graphicx}
\usepackage{breakurl}

\doi{10.1214/13-AOAS670} 
\volume{7}
\issue{4}
\pubyear{2013}
\firstpage{2034}
\lastpage{2061}

\makeatletter
\def\rmfamily{}
\newcolumntype{d}[1]{D{.}{.}{#1}}

\renewcommand{\citep}[1]{(\citeauthor{#1} \citeyear{#1})}
\makeatother

\begin{document}
\begin{frontmatter}

\title{Bayesian spline method for assessing extreme loads on wind turbines}
\runtitle{Spline method for extreme loads}

\begin{aug}
\author[a]{\fnms{Giwhyun} \snm{Lee}\thanksref{m1,t1}\ead[label=e1]{giwhyunlee@gmail.com}},
\author[b]{\fnms{Eunshin} \snm{Byon}\thanksref{m2,t1}\ead[label=e2]{ebyon@umich.edu}},
\author[a]{\fnms{Lewis} \snm{Ntaimo}\thanksref{m1}\ead[label=e3]{ntaimo@tamu.edu}}
\and\break
\author[a]{\fnms{Yu} \snm{Ding}\thanksref{m1,t1}\corref{}\ead[label=e4]{yuding@iemail.tamu.edu}}
\thankstext{t1}{Supported in part by NSF Grants CMMI-0727305 and CMMI-0926803.}
\runauthor{Lee, Byon, Ntaimo and Ding}
\affiliation{Texas A\&M University\thanksmark{m1} and University of
Michigan\thanksmark{m2}}
\address[a]{G. Lee\\
L. Ntaimo\\
Y. Ding\\
Department of Industrial and\\
\quad Systems Engineering \\
Texas A\&M University\\
College Station, Texas 77843-3131\\
USA\\
\printead{e1}\\
\phantom{E-mail:\ }\printead*{e3}\\
\phantom{E-mail:\ }\printead*{e4}}

\address[b]{E. Byon\\
Department of Industrial and \\
\quad Operations Engineering \\
University of Michigan\\
Ann Arbor, Michigan 48109\\
USA\\
\printead{e2}}
\end{aug}

\received{\smonth{9} \syear{2012}}
\revised{\smonth{6} \syear{2013}}

%
\begin{abstract}
This study presents a Bayesian parametric model for the purpose of
estimating the extreme load on a wind turbine. The extreme load is the
highest stress level imposed on a turbine structure that the turbine
would experience during its service lifetime. A wind turbine should be
designed to resist such a high load to avoid catastrophic structural
failures. To assess the extreme load, turbine structural responses are
evaluated by conducting field measurement campaigns or performing
aeroelastic simulation studies. In general, data obtained in either
case are not sufficient to represent various loading responses under
all possible weather conditions. An appropriate extrapolation is
necessary to characterize the structural loads in a turbine's service
life. This study devises a Bayesian spline method for this
extrapolation purpose, using load data collected in a period much
shorter than a turbine's service life. The spline method is applied to
three sets of turbine's load response data to estimate the
corresponding extreme loads at the roots of the turbine blades.
Compared to the current industry practice, the spline method appears to
provide better extreme load assessment.
\end{abstract}

%

\end{frontmatter}

\section{Introduction}\label{secintro}
A wind turbine operates under various loading conditions in stochastic
weather environments. The increasing size, weight and length of
components of utility-scale wind turbines escalate the stresses (or
loads, responses) imposed on the structure. As a result, modern wind
turbines are prone to experiencing structural failures. Of particular
interest in a wind turbine system are the extreme events under which
loads exceed a threshold, called a ``nominal design load'' or ``extreme
load.'' Upon the occurrence of a load higher than the nominal design
load, a wind turbine could experience catastrophic structural failures.

Mathematically, an extreme load is defined as an extreme quantile value
in a load distribution corresponding to a turbine's service time of $T$
years [\citet{SN2007}].
Let $y$ denote the maximum load, in the unit of million Newton-meter
(MN-m), during a specific time interval. Then, we define the load
exceedance probability as follows:
%
%
\begin{equation}
P_T=P[y>l_T] \label{eqnlong},
\end{equation}
where $P_T$ is the target probability of exceeding the load
level $l_T$ (in the same unit as that of $y$).

In structural reliability analysis of wind turbines, people collect
load response data and arrange them in 10-minute intervals because wind
speeds are considered stationary over a 10-minute duration [\citet
{FW2001}]. Given this data arrangement in wind industry, $y$ commonly
denotes the maximum load during a 10-minute interval. The unconditional
distribution of $y$, $p(y)$, is called the \textit{long-term}
distribution and is used to calculate $P[y>l_T]$ in~(\ref{eqnlong}).

In~(\ref{eqnlong}), the extreme event, $ \{y>l_T \}$, takes
place with the exceedance probability $P_T$. The waiting time until
this event happens should be longer than, or equal to, the service
time. Therefore, a reasonable level of $P_T$ can be found in the
following way [\citet{IEC,Peeringa2003}]:
%
%
\begin{equation}
P_T = \frac{10}{T \times365.25 \times24 \times60}. \label{eqnpt}
\end{equation}
Note that $P_T$ is the reciprocal of the number of 10-minute
intervals in $T$ years. For example, when $T$ is $50$, $P_T$ becomes
$3.8 \times10^{-7}$.

Estimating the extreme load implies finding an extreme quantile $l_T$
in the 10-minute maximum load distribution, given a target service
period $T$, such that (\ref{eqnlong}) is satisfied. Wind turbines
should be designed to resist the $l_T$ load level to avoid structural
failures during its desired service life.

Since loads are highly affected by wind profiles, we consider the
marginal distribution of $y$ obtained by using the distribution of $y$
conditional on a wind profile as follows:
%
%
\begin{equation}
p(y)=\int p(y|\mathbf{x})p(\mathbf{x}) \,d\mathbf{x}. \label{eqnshort}
\end{equation}
Here, $p(\mathbf{x})$ is the joint probability density
function of wind characteristics in a covariate vector $\mathbf{x}$.
The conditional distribution of $y$ given $\mathbf{x}$,
$p(y|\mathbf{x})$ in~(\ref{eqnshort}), is called the \textit
{short-term} distribution. The long-term distribution can be computed
by integrating out wind characteristics in the short-term distribution.

The conditional distribution modeling in (\ref{eqnshort}) is a
necessary practice in the wind industry. A turbine needs to be assessed
for its ability to resist the extreme loads under the specific
wind\vadjust{\goodbreak}
profile at the site it will be installed. Turbine manufacturers usually
test a small number of representative turbines at their own testing
site, producing $p(y|\mathbf{x})$. When a turbine is to be installed
at a commercial wind farm, the wind profile at the proposed
installation site can be collected and substituted into~(\ref
{eqnshort}) as $p(\mathbf{x})$, so that the site-specific extreme
load can be assessed. Without the conditional distribution model, a
turbine test would have to be done for virtually every new wind farm;
doing so is very costly and thus uncommon.

For in-land turbines, the wind characteristic vector $\mathbf{x}$ in
general comprises two elements: (1) a steady state mean of wind speed
and (2) the stochastic variability of wind speed [\citet
{BCC2010,RL2000,MVW2001}]. The first element can be measured by the
average wind speed (in the unit of meters per second, or m$/$s) during a
10-minute interval, and the second element can be represented by the
standard deviation of wind speed, or the turbulence intensity, also
during a 10-minute interval. Here, turbulence intensity is defined as
the standard deviation of wind speed divided by the average wind speed
for the same duration. For offshore turbines, weather characteristics
other than wind may be needed, such as the wave height [\citet{AM2008}].

In this study, we propose a new procedure to estimate the long-term
extreme load level $l_T$ for wind turbines. The novelty of the new
procedure is primarily regarding how to model the short-term
distribution $p(y|\mathbf{x})$. Specially, we establish a load
distribution for $y|\mathbf{x}$ using spline models. As such, we
label the resulting method \emph{a Bayesian spline method for extreme
loads}. In the remainder of the paper we first provide some background
information regarding wind turbine load responses and the data sets
used in this study. In Section~\ref{secreview} we explain how the
extreme load estimation problem is currently solved. We proceed to
present the details of our spline method in Section~\ref{secapproach}.
In Section~\ref{secresult} we compare the spline method with the
method reviewed in Section~\ref{secreview}, arguing that the spline
method produces better estimates. Finally, we end the paper with some
concluding remarks in Section~\ref{secsummary}.

\section{Background and data sets}\label{secbkdata}

\begin{figure}

\includegraphics{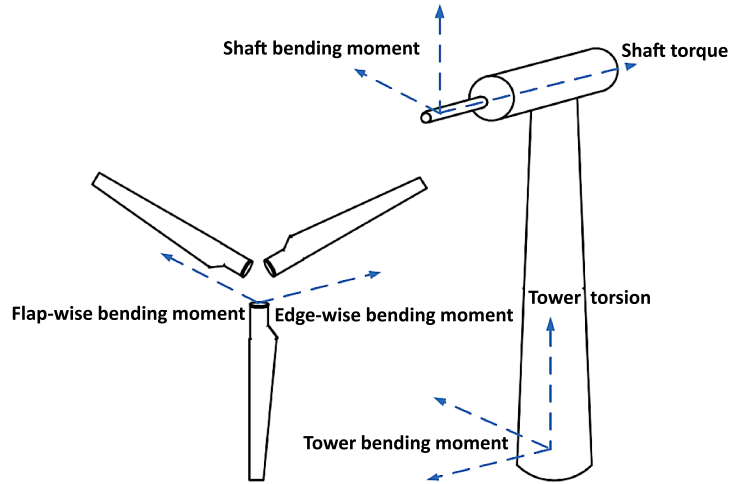}

\caption{Illustration of structural loads at different
components. (The illustration is modified based on a
figure originally available at \citeauthor{website1}.)}\label{figturbine}
\end{figure}

Figure~\ref{figturbine} shows examples of mechanical loads at different components in
a turbine system. The flap-wise bending moments measure the loads at
the blade roots that are perpendicular to the rotor plane, while the
edge-wise bending moments measure the loads that are parallel to the
plane. Shaft- and tower-bending moments measure, in two directions,
the stresses on the main shaft connected to the rotor and on the tower
supporting the wind power generation system (i.e., blades, rotor,
generator etc.), respectively.
%

We only study in-land turbines (ILTs) in this work and use the data
sets from three ILTs located at different sites. These data sets were
collected by Ris{\o}-DTU (Technical University of Denmark) [\citeauthor
{website1}]. Table~\ref{tblturbine} summarizes the specification of
the data sets.\vadjust{\goodbreak}

\begin{table}[b]
\caption{Specifications of wind turbines in three data sets} \label{tblturbine}
\begin{tabular*}{\textwidth}{@{\extracolsep{\fill}}lccc@{}}
\hline
\textbf{Wind turbine model} & \textbf{NEG-Micon/2750} &
\textbf{Vestas V39} & \textbf{Nordtank 500} \\
\textbf{(Name of data set)}& \textbf{(ILT1)} & \textbf{(ILT2)} &
\textbf{(ILT3)}\\
\hline
Hub height (m) & 80 & 40 & 35\\
Rotor diameter (m) & 92 & 39 & 41\\
{Cut-in} wind speed (m$/$s) & 4 & 4.5 & 3.5\\
{Cut-out}
wind speed (m$/$s) & 25 & 25 & 25\\
 Rated wind speed (m$/$s)
& 14 & 16 & 12 \\
 Nominal power (kW) & 2750 & 500 & 500\\
 Control system & Pitch & Pitch & Stall\\
Location & Alborg, & Tehachapi Pass, & Roskilde,\\
&
Denmark & California & Denmark\\ Terrain & Coastal &
Bushes & Coastal\\
\hline
\end{tabular*}
\end{table}

We would like to first explain a few terms used in the table as well as
in the rest of the paper:
\begin{itemize}
\item\emph{Pitch control}: To avoid production of excessive
electricity, turbines hold the rotor at an approximately constant speed
in high wind speeds. A pitch controlled turbine turns its blades to
regulate its rotor speed.
\item\emph{Stall control}: This serves the same purpose as in pitch
control. But the blade angles do not adjust during operation. Instead
the blades are designed and shaped to increasingly stall the blade's
angle of attack with the wind to protect the turbine from excessive
wind speeds.
\item\emph{Cut-in wind speed}: This is the lowest wind speed at a hub
height at which a wind turbine starts to produce power.
\item\emph{Cut-out wind speed}: This is the speed beyond which a wind
turbine shuts itself down to protect the turbine.
\item\emph{Rated wind speed}: This is the speed beyond which the
turbine's output power needs to be limited and, consequently, the rotor
speeds are regulated, by using, for example, a pitch control mechanism.
\end{itemize}

Among the structural load responses, we consider only the flap-wise
bending moments measured at the root of blades. In other words, $y$ in
this study is the 10-minute maximum blade-root flap-wise bending moment
(hereafter, we call $y$ a maximum load). But please note that our
method applies to other load responses as well. Regarding weather
characteristics, since we consider only the ILTs, we include in
$\mathbf{x}$ the average wind speed $v$ and the standard deviation
of wind speed $s$, namely, $\mathbf{x}:=(v,s)$.

The data are recorded at different frequencies on the ILTs, as follows:
\begin{itemize}
\item\emph{ILT1}: 25 Hz${} = {}$15,000 measurements/10-min;
\item\emph{ILT2}: 32 Hz${} = {}$19,200 measurements/10-min;
\item\emph{ILT3}: 35.7 Hz${} = {}$21,420 measurements/10-min.
\end{itemize}
Here, 1 Hz means one measurement per second. The raw measured variables
are $v_{ij}$ and $y_{ij}$, where $i =1,\ldots,n$ represents a 10-minute
block of data and $j =1,\ldots,N$ is the index of the measurements. We
use $N$ to represent the number of measurements in a 10-minute block,
equal to 15,000, 19,200 and 21,420 for ILT1, ILT2 and ILT3,
respectively, and use $n$ to represent the total number of the
10-minute intervals in each data set, taking the value of 1154, 595
and 5688, respectively, for ILT1, ILT2 and ILT3. For these variables,
the statistics of the observations in each 10-minute block are
calculated as follows:
%
\begin{eqnarray}
v_i&=&\frac{1} {N} \sum_{j=1}^{N}
v_{ij}, \label{eqn21}
\\
s_i&=&\sqrt{\frac{1} {N-1} \sum
_{j=1}^{N} (v_{ij}-v_i)^2}
\quad \mbox{and}\label {eqn22}
\\
y_i&=&\max \{ y_{i1},y_{i2},
\ldots,y_{iN} \}. \label{eqn23}
\end{eqnarray}

\section{Literature review} \label{secreview}

The previous edition of the international standard, IEC 61400-1:1999,
offers a set of design load cases with \emph{deterministic} wind
conditions such as annual average wind speeds, higher and lower
turbulence intensities, and extreme wind speeds [\citet{IEC99}]. In
other words, the loads in IEC 61400-1:1999 are specified as discrete
events based on design experiences and empirical models [\citet
{MHB2002}]. \citet{VB2001} point out that these deterministic models do
not represent the stochastic nature of structure responses, and suggest
using statistical modeling to improve design load estimates. \citet
{MHB2002} examine the effect of varying turbulence levels on the
statistical behavior of a wind turbine's extreme load. They conclude
that the loading on a turbine is stochastic at high turbulence levels,
significantly influencing the tail of the load distribution.

In response to these developments, the new edition of IEC 61400-1
standard (IEC 61400-1:2005), issued in 2005, replaces the deterministic
load cases with \emph{stochastic} models, and recommends the use of
\emph{statistical} approaches for determining the extreme load level in
the design stage. \citet{FA2008} compare the deterministic load cases in
the IEC 61400-1:1999 with the stochastic cases in IEC 61400-1:2005, and
observe that when statistical approaches are applied, higher extreme
load estimates are obtained in some structural responses, such as the
blade tip deflection and flap-wise bending moment.

After IEC 61400-1:2005 was issued, many studies were reported to devise
and recommend statistical approaches for extreme load analysis [\citet
{FA2008,AM2008,Peeringa2009,Moriarty2008,FAM2008,RM2008,NH2008}].
These studies adopt a common framework, which we call \emph{binning}
method. The basic idea of the binning method is to discretize the
domain of a wind profile vector $\mathbf{x}$ into a finite number of
bins. For example, one can divide the range of wind speed, from the
\emph{cut-in} speed to the \emph{cut-out} speed, into multiple bins and
set the width of each bin to, say, 2 m$/$s. Then, in each bin, the
conditional short-term distribution of $y|\mathbf{x}$ is
approximated by a stationary distribution, with the parameters of the
distribution estimated by the method of moments or the maximum
likelihood method. Then, the contribution from each bin is summed over
all possible bins to determine the final long-term extreme load. In
other words, integration in (\ref{eqnshort}) for calculating the
long-term distribution is approximated by the summation of finite elements.

According to the classical extreme value theory [\citet{Coles2001,Smith1990}], the short-term distribution of $y|\mathbf{x}$ can be
approximated by a generalized extreme value (GEV) distribution. The
probability density function of the GEV is
%
%
\begin{equation}
\label{eqngev}\qquad p(y)= %
\cases{\displaystyle \frac{1}{\sigma}\exp \biggl[ -
\biggl(1+\xi \biggl(\frac{y-\mu
}{\sigma} \biggr) \biggr)^{-{1}/{\xi}} \biggr]
\biggl(1+\xi \biggl(\frac
{y-\mu}{\sigma} \biggr) \biggr)^{-1-{1}/{\xi}},\vspace*{2pt}\cr
\hspace*{159pt}\qquad \mbox{if } \xi \neq0, \vspace*{2pt}
\cr
\displaystyle\frac{1}{\sigma}\exp \biggl[-
\frac{y-\mu}{\sigma} - \exp \biggl(-\frac
{y-\mu}{\sigma} \biggr) \biggr],\qquad \mbox{if } \xi= 0,} %
\end{equation}
for $\{y \dvtx  1+\xi(y-\mu)/\sigma> 0 \}$, where $\mu\in\Re$
is the location parameter, $\sigma>0$ is the scale parameter, and $\xi
\in\Re$ is the shape parameter that determines the weight of the tail
of the distribution. $\xi>0$ corresponds to the Fr\'{e}chet
distribution with a heavy upper tail, $\xi<0$ to the Weibull
distribution with a short upper tail and light lower tail, and $\xi=0$
(or, $\xi\rightarrow0$) to the Gumbel distribution with a light upper
tail [\citet{Coles2001}].

One of the main focuses of interest in extreme value theory is in
deriving the quantile value (which, in our study, is defined as the
extreme load level~$l_T$), given the target probability $P_T$. The
quantile value can be expressed as a function of the distribution
parameters as follows:
%
%
\begin{eqnarray}
\label{eqnquan} l_T = %
\cases{\displaystyle \mu-\frac{\sigma}{\xi}
\bigl[1- \bigl(-\log (1-P_T ) \bigr)^{-\xi} \bigr], &\quad $
\mbox{if } \xi\neq0$, \vspace*{2pt}
\cr
\mu-\sigma\log \bigl[-\log
(1-P_T ) \bigr], &\quad $\mbox{if } \xi= 0.$} %
\end{eqnarray}

The virtue of the binning method is that by modeling the short-term
distribution with a homogeneous GEV distribution (i.e., keep the
parameters therein constant), it provides a simple way to handle the
overall nonstationary load response across different wind speeds. The
binning method is perhaps the most common method used in the wind
industry and also recommended by \citet{IEC}. For example, \citet{AM2008}
use the binning method to estimate the extreme loads for a 2MW offshore
wind turbine. In each weather bin, they use the Gumbel distribution to
explain the probabilistic behavior of the mudline bending moments of
the turbine tower. The data were collected for a period of 16 months.
However, most bins have a small number of data, or sometimes, no data
at all. For the bins without data, the authors estimate the short-term
distribution parameters by using a weighted average of all nonempty
bins with the weight related to the inverse squared distance between
bins. They quantify the uncertainty of the estimated extreme loads
using a bootstrapping technique and report 95\% confidence intervals
for the short-term extreme load given specific weather conditions
(weather bins). Because bootstrapping resamples the existing data for a
given weather bin, it cannot precisely capture the uncertainty for
those bins with limited data or without data.

Despite its popularity, the binning method has obvious shortcomings in
estimating extreme loads. A major limitation is that the short-term
load distribution in one bin is constructed separately from the
short-term distributions in other bins. This approach requires an
enormous amount of data to define the tail of each short-term
distribution. In reality, the field data can only be collected in a
short duration (e.g., one year out of the 50-year service) and,
consequently, some bins do not have enough data. Then, the binning
method may end up with inaccuracy or big uncertainty in the estimates
of extreme loads. In practice, how many bins to use is also under
debate, and there is not yet a consensus. The answer to the action of
binning appears to depend on the amount of data---if one has more
data, he/she can afford to use more bins; otherwise, fewer bins.

\section{Bayesian spline method for extreme load}\label{secapproach}
In this section we present our new procedure of estimating the extreme
load with two submodels. 
The first submodel (in Section~\ref{subsecspline}) is the conditional
maximum load model $p(y|\mathbf{x})$, and the second submodel (in
Section~\ref{subsecenv}) is the distribution of wind characteristics
$p(\mathbf{x})$. Our major undertaking in this study is on the first
submodel, where we present an alternative to the current binning
method. 

\begin{figure}

\includegraphics{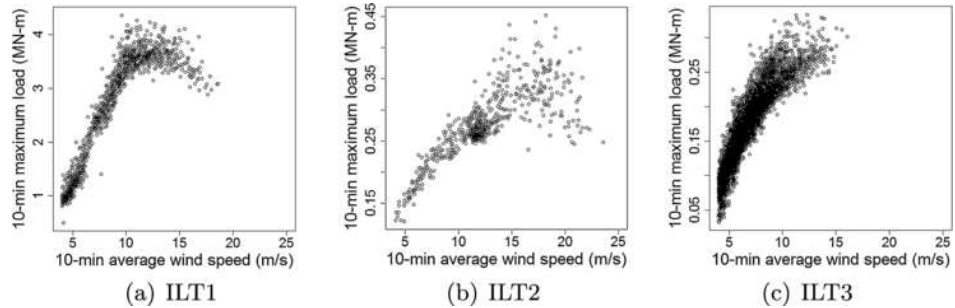}

\caption{Scatter plots of 10-minute maximum load
versus 10-minute average wind speed.} \label{figSC1}
\end{figure}

We begin by presenting some scatter plots for the three data sets.
Figure~\ref{figSC1} shows the scatter plots between the 10-minute
maximum loads and 10-minute average wind speeds. We observe nonlinear
patterns between the loads and the average wind speeds in all three
scatter plots, while individual turbines exhibit different response
patterns. ILT1 and ILT2 are two pitch controlled turbines, so when the
wind speed reaches or exceeds the rated speed, the blades are adjusted
to reduce the absorption of wind energy. As a result, we observe that
the loads show a downward trend after the rated wind speed. But
different from that of ILT1, the load response of ILT2 has a large
variation beyond the rated wind speed. This large variation can be
attributed to its less capable control system since ILT2 is one of the
early turbine models using a pitch control system. ILT3 is a stall
controlled turbine, and its load pattern in Figure~\ref{figSC1}(c)
does not have an obvious downward trend beyond the rated speed.

\begin{figure}

\includegraphics{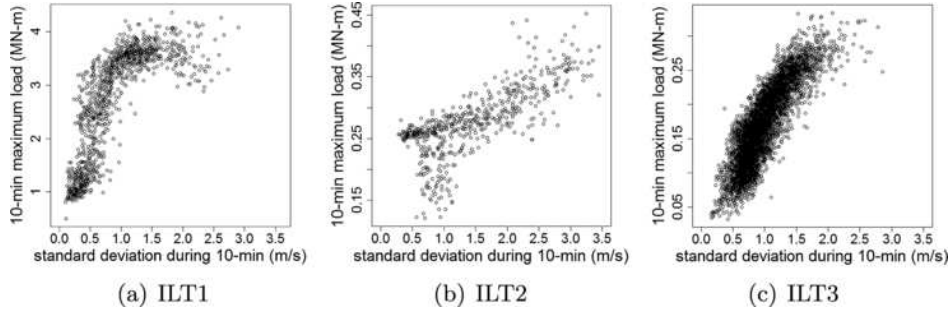}

\caption{Scatter plots of 10-minute maximum load
versus 10-minute standard deviations of wind speed.} \label{figSC2}
\end{figure}

\begin{figure}[b]

\includegraphics{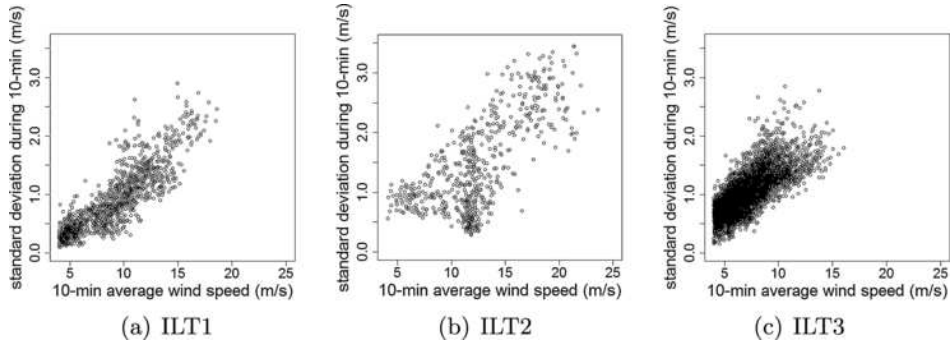}

\caption{Scatter plots of 10-minute average wind
speed versus 10-minute standard deviation of wind speed.} \label{figSC3}
\end{figure}

Figure~\ref{figSC2} presents the scatter plots between the 10-minute
maximum loads and the standard deviations of wind speed during the
10-minute intervals. We also observe nonlinear relationships between
them, especially for the new pitch-controlled ILT1. Figure~\ref{figSC3} shows scatter plots of 10-minute standard deviation versus
10-minute average wind speed. Some previous studies [\citet{MHB2002,FCV2003}] suggest that the standard deviation of wind speed varies with
the average wind speed, which appears consistent with what we observe
in Figure~\ref{figSC3}.

%

\subsection{Submodel 1: Bayesian spline model for conditional maximum
load}\label{subsecspline}
Recall that in the binning method, a homogeneous GEV distribution is
used to model the short-term load distribution, for it appears
reasonable to assume stationarity if the chosen weather bin is narrow
enough. A finite number of the homogeneous GEV distributions are then
stitched together to represent the nonstationary nature across the
entire wind profile. What we propose here is to abandon the bins and
instead use a nonhomogeneous GEV distribution whose parameters are not
constant but depend on weather conditions.

Our research started out with simple approaches based on polynomial
models. It turns out that polynomial-based approaches lack the
flexibility of adapting to the data sets from different types of
turbines. Moreover, due to the nonlinearity around the rated wind speed
and the limited amount of data under high wind speeds, polynomial-based
approaches performed poorly in those regions that are generally
important for capturing the maximum load. Spline models, on the other
hand, appear to work better than a global polynomial model, because
they have more supporting points spreading over the input regions. In
the sequel, we present two flexible Bayesian spline models for the
purpose of establishing the desired nonhomogeneous GEV distribution.

Suppose we observe 10-minute maximum loads $y_1, \ldots, y_n$ with
corresponding covariate variables $\mathbf{x}_1 =(v_1,s_1), \ldots
, \mathbf{x}_n = (v_n, s_n)$, as defined in (\ref
{eqn21}) and (\ref{eqn22}). We choose to model $y_i$ with a GEV distribution:
%
%
\begin{equation}
\label{eqngevmodel} y_i | \mathbf{x}_i \sim \operatorname{GEV}
\bigl( \mu(\mathbf{x}_i), \sigma (\mathbf{x}_i),  \xi
\bigr), \qquad\sigma(\cdot)>0,
\end{equation}
where the location parameter $\mu$ and scale parameter $\sigma$ in this
GEV distribution are a nonlinear function of wind characteristics
$\mathbf{x}$. The shape parameter $\xi$ is fixed across the wind
profile, while its value will still be estimated using the data from a
specific wind turbine. The reason that we keep $\xi$ fixed is to keep
the final model from becoming overly complicated. Let us denote $\mu
(\mathbf{x}_i)$ and $\sigma(\mathbf{x}_i)$ by
%
%
\begin{eqnarray}
\mu(\mathbf{x}_i) &=&f(\mathbf{x}_i), \label{eqnmu}
\\
\sigma(\mathbf{x}_i) &=& \exp \bigl(g(\mathbf{x}_i)
\bigr),\label{eqnsigma}
\end{eqnarray}
where in (\ref{eqnsigma}), an exponential function is used to ensure
the positivity of the scale parameter.

Our strategy of modeling $f(\cdot)$ and $g(\cdot)$ is to use a Bayesian
MARS (multivariate adaptive regression splines) model [\citet{DMS19981,BHMS2002}] for capturing the nonlinearity between the load response and
the wind-related covariates. The Bayesian MARS model has high
flexibility. It includes the number and locations of knots as part of
its model parameters and determines these from observed data. In
addition, interaction effects among input factors can be modeled if
choosing appropriate basis functions.

Specifically, the Bayesian MARS models $f(\mathbf{x})$ for the
location parameter $\mu$ and $g(\mathbf{x})$ for the scale
parameter $\sigma$ are represented as a linear combination of the basis
functions $B^\mu_{k}(\mathbf{x})$ and $B^\sigma_{k}(\mathbf{x})$, respectively, as
%
%
\begin{eqnarray}
f(\mathbf{x}) &=& \sum_{k=1} ^{K_\mu}
\beta_{k} B^\mu_k(\mathbf{x}), \label{eqnmu1}
\\
g(\mathbf{x}) &=& \sum_{k=1} ^{K_\sigma}
\theta_{k} B^\sigma _{k}(\mathbf{x}),
\label{eqnsigma1}
\end{eqnarray}
where $\beta_{k}, k=1,\ldots,K_\mu$ and $\theta_{k}, k=1,\ldots,K_\sigma
$ are the coefficients of the basis functions $B^\mu_{k}(\cdot)$ and
$B^\sigma_{k}(\cdot)$, respectively, and $K_\mu$ and $K_\sigma$ are the
number of the respective basis functions. According to the study by
\citet{DMS19981}, which proposed the Bayesian MARS, the basis functions
are specified as follows:
%
%
\begin{equation}\label{eqnmu2}
B_k(\mathbf{x}) = \cases{ %
1,&\quad $k=1$,
\vspace*{2pt}\cr
\displaystyle\prod_{j=1} ^{J_k} \bigl[h_{jk}
\cdot(x_{r(j,k)} - t_{jk}) \bigr]_{+},&\quad $k=2,3,\ldots,
K$. }
\end{equation}
Here, $[\cdot]_{+}=\max(0,\cdot)$, $J_k$ is the degree of
interaction modeled by the basis function $B_k(\mathbf{x})$,
$h_{jk}$ is the sign indicator, taking the value of either $-1$ or
$+1$, and $r(j,k)$ produces the index of the predictor variable which
is being split on $t_{jk}$, commonly referred to as the knot points.

We here introduce an integer variable $T_k$ to represent the types of
basis functions used in (\ref{eqnmu2}). Since we consider two
predictors $v$ and $s$ for inland turbines, there could be three types
of basis functions, namely, $[\pm(v - *)]_{+}$ and $[\pm(s - *)]_{+}$
for each explanatory variable, respectively, and $[\pm(v - *)]_{+}[\pm
(s - *)]_{+}$ for interactions between them. So we let $T_k$ take
the integer value of 1, 2 or 3, to represent the three types of basis
functions. That is, $[\pm(v - *)]_{+}$ is represented by $T_k =1$, $[\pm
(s - *)]_{+}$ represented by $T_k=2$, and $[\pm(v - *)]_{+}[\pm(s -
*)]_{+}$ represented by $T_k=3$. When $J_k =1$ in equation (\ref
{eqnmu2}), then the first two types of basis functions are used, while
when $J_k=2$, all three types of basis functions are used. In our
model, we set $J_k = 1$ or $J_k = 2$ in the model of the location
parameter $\mu$ for ILT1 and ILT3 data to allow the interaction to be
modeled. For ILT2, however, due to its relatively smaller data amount,
a model setting $J_k = 2$ produces unstable and unreasonably wide
credible intervals. So for ILT2, $J_k =1 $ is set for its location
parameter $\mu$. For the scale parameter $\sigma$, we set $J_k=1$ for
all three data sets, but for ILT2, again due to its data scarcity, we
include $v$ as the only predictor in its scale parameter model.


Let $\bolds{\Psi}_a = (\bolds{\Psi}_{\mu},\bolds{\Psi
}_{\sigma},\xi)$ denote all the parameters used in model (\ref
{eqngevmodel}), where $\bolds{\Psi}_{\mu}$ and $\bolds{\Psi
}_{\sigma}$ include the parameters in function $f(\cdot)$ and $g(\cdot
)$, respectively. These parameters are grouped into two sets: (1) the
coefficients of the basis functions in $\bolds{\beta}=(\beta
_1,\ldots,\beta_{K_\mu})$ or $\bolds{\theta}=(\theta_1,\ldots
,\theta_{K_\sigma})$, and (2) the number and locations of the knots,
and the types of basis function in $\bolds{\phi}_{\mu}$ or
$\bolds{\phi}_{\sigma}$, as follows:
%
%
\begin{eqnarray}
\label{eqnphimu} \bolds{\phi}_{\mu}= \bigl(K_\mu, \bolds{
\Lambda}^\mu_2, \ldots, \bolds{\Lambda}^\mu_{K_\mu}
\bigr),
\nonumber
\\[-8pt]
\\[-8pt]
\eqntext{\mbox{where } \bolds{\Lambda}^\mu_{k} = \cases{
\bigl(T^\mu_{k},
h^\mu_{1k}, t^\mu_{1k} \bigr),&\quad
$\mbox{when } T^\mu_k=1, 2$;
\vspace*{2pt}\cr
\bigl(T^\mu_{k}, h^\mu_{1k},
h^\mu_{2k},t^\mu_{1k},
t^\mu_{2k} \bigr), &\quad $\mbox{when } T^\mu_k=3,$}}
\end{eqnarray}
and
%
%
\begin{eqnarray}
\label{eqnphisigma} \bolds{\phi}_{\sigma}&=& \bigl(K_\sigma,
\bolds{ \Lambda}^\sigma _2, \ldots, \bolds{
\Lambda}^\sigma_{K_\sigma} \bigr),
\nonumber
\\[-8pt]
\\[-8pt]
\eqntext{\mbox{where } \bolds{\Lambda}^\sigma_{k} =
\bigl(T^\sigma_{k}, h^\sigma_{1k},
t^\sigma_{1k} \bigr) \mbox{ when } T^\sigma
_k=1,2.}
\end{eqnarray}
Using the above notation, we have $\bolds{\Psi}_{\mu} =
(\bolds{\beta}, \bolds{\phi}_{\mu})$ and $\bolds{\Psi
}_{\sigma} = (\bolds{\theta}, \bolds{\phi}_{\sigma})$.\vadjust{\goodbreak}

To complete the Bayesian formulation for the model in (\ref
{eqngevmodel}), priors of the parameters involved should be specified.
In this paper, we use uniform priors on $\bolds{\phi}_\mu$ and
$\bolds{\phi}_\sigma$; see the detailed expression in Appendix~\ref
{secPRI}. Given $\bolds{\phi}_{\mu}$ and $\bolds{\phi
}_{\sigma}$, we specify the prior distribution for the parameters
$(\bolds{\beta},\bolds{\theta},\xi)$ as the unit-information
prior, that is, UIP [\citet{KW1995}], which is defined by setting the
corresponding covariance matrix to be equal to the Fisher information
of one observation. 

\subsection{Submodel 1: Posterior distribution of parameters}\label
{subsecposterior}
The Bayesian\break  MARS model treats the number and locations of the knots as
random quantities. When the number of knots changes, the dimension of
the parameter space changes with it. To handle a varying dimensionality
in the probability distributions in a random sampling procedure,
researchers usually use a reversible jump Markov chain Monte Carlo
(RJMCMC) algorithm developed by \citet{Green1995}. The acceptance
probability for a RJMCMC algorithm includes a Jacobian term, which
accounts for the change in dimension. However, under the assumption
that the model space for parameters of varying dimension is discrete,
there is no need for a Jacobian. In our analysis, this assumption is
satisfied since we only consider probable models over all possible knot
locations and numbers. Therefore, instead of using the RJMCMC
algorithm, we use the reversible jump sampler (RJS) algorithm proposed
in \citet{BHMS2002}. Since the RJS algorithm does not require new
parameters to match dimensions between models and the corresponding
Jacobian term to the acceptance probability, it is simpler and more
efficient to execute.

To allow for dimensional changes, there are three actions in the RJS
algorithm: \rmfamily{BIRTH}, \rmfamily{DEATH} and \rmfamily{MOVE},
which adds, deletes or alters a basis function, respectively.
Accordingly, the number of knots as well as the locations of some knots
change. The detailed definitions of the three actions are given in
\citet{BHMS2002}, page 53, so we need not repeat them here. They
suggest the following: use equal probability (i.e., $\frac{1}{3}$) to
propose any of the three moves, and then use the following acceptance
probability $\alpha$ for a proposed move from a model having $k$ basis
functions to a model having $k^c$ basis functions:
%
%
\begin{equation}
\alpha= \min \{1, \mbox{the ratio of marginal likelihood } \times R \},
\label{eqnaccmod1}
\end{equation}
where $R$ is a ratio of probabilities defined as follows:
\begin{itemize}
\item For a \rmfamily{BIRTH} action, $R = \frac{\mbox{probability of
DEATH in model $k^{c}$}}{\mbox{probability of BIRTH in model $k$}}$;
\item For a \rmfamily{DEATH} action, $R = \frac{\mbox{probability of
BIRTH in model $k^{c}$}}{\mbox{probability of DEATH in model $k$}}$;
\item For a \rmfamily{MOVE} action, $R = \frac{\mbox{probability of
MOVE in model $k^{c}$}}{\mbox{probability of MOVE in model $k$}}$.
\end{itemize}
We have $R=1$ for most cases, because the probabilities in the
denominator and numerator are equal, except when $k$ reaches either the
upper or the lower bound.

The marginal likelihood in (\ref{eqnaccmod1}) can be expressed as follows:
%
\begin{eqnarray}
\label{eqnmarg} &&p (\mathcal{D}_y|\bolds{\phi}_{\mu},
\bolds{ \phi}_{\sigma
} )
\nonumber
\\[-8pt]
\\[-8pt]
\nonumber
&&\qquad=
\int{p (\mathcal{D}_y|\bolds{\beta },\bolds{ \theta},\xi,\bolds{
\phi}_{\mu},\bolds{\phi }_{\sigma} )p (\bolds{ \beta},\bolds{
\theta}, \xi |\bolds{\phi}_{\mu}, \bolds{ \phi}_{\sigma} )\,d
\bolds{\beta} \,d\bolds{\theta} \,d\xi},
\end{eqnarray}
where $\mathcal{D}_y= ({y_1, \ldots, y_n} )$
represents a set of observed load data. Since it is difficult to
calculate the above marginal likelihood analytically in our study, we
consider an approximation of $p  (\mathcal{D}_y|\bolds{\phi
}_{\mu}, \bolds{\phi}_{\sigma} ) $. \citet{KW1995} and \citet
{Raftery1995} showed that when UIP priors are used, the marginal
log-likelihood, that is, $\log (p (\mathcal{D}_y|\bolds{\phi}_{\mu}, \bolds{\phi}_{\sigma} ) )$, can be
reasonably approximated by the Schwarz information criterion (SIC)
[\citet{Schwarz1978}]. The SIC is expressed as
\[
\label{eqnSIC} \mathrm{SIC}_{\bolds{\phi}_\mu, \bolds{\phi}_\sigma} = \log \bigl(p(\mathcal{D}_y|
\hat{ \bolds{\beta}},\hat{\bolds{\theta }},\hat{\xi},\bolds{\phi}_{\mu},
\bolds{ \phi}_{\sigma}) \bigr)- \tfrac{1}{2}\,d_{k}\log(n)
,
\]
where $\hat{\bolds{\beta}}, \hat{\bolds{\theta}},
\hat{\xi}$ are the maximum likelihood estimators (MLEs) of the
corresponding parameters obtained conditional on $\bolds{\phi}_{\mu
}$ and $\bolds{\phi}_{\sigma}$, and $d_k$ is the total number of
parameters to be estimated. In this case, $d_k=K_\mu+ K_\sigma+ 1$.

Recall that we have two dimension-varying states $\bolds{\phi}_\mu
$ and $\bolds{\phi}_\sigma$ in the RJS algorithm. Depending on
which state vector is changing, two marginal log-likelihood ratios are
needed, and they are approximated by the corresponding SICs, such as
%
%
\begin{eqnarray}
\log\frac{p  (\mathcal{D}_y|\bolds{\phi}_{\mu
}^c, \bolds{\phi}_{\sigma} )}{ p  (\mathcal
{D}_y|\bolds{\phi}_{\mu}, \bolds{\phi}_{\sigma} )} &\backsimeq&\mathrm{SIC}_{\bolds{\phi}^c_\mu, \bolds{\phi
}_\sigma}-
\mathrm{SIC}_{\bolds{\phi}_\mu, \bolds{\phi}_\sigma}\quad \mbox{and}
\\
\log\frac{p  (\mathcal{D}_y|\bolds{\phi}_{\mu
}, \bolds{\phi}_{\sigma}^c )}{ p  (\mathcal
{D}_y|\bolds{\phi}_{\mu}, \bolds{\phi}_{\sigma} )} &\backsimeq&\mathrm{SIC}_{ \bolds{\phi}_\mu,\bolds{\phi
}^c_\sigma}-
\mathrm{SIC}_{\bolds{\phi}_\mu,\bolds{\phi}_\sigma}.
\end{eqnarray}

Then, we use two acceptance probabilities $\alpha_\mu$ and $\alpha
_\sigma$ for accepting or rejecting a new state in $\bolds{\phi
}_\mu$ and $\bolds{\phi}_\sigma$, respectively. Using the SICs,
$\alpha_\mu$ and $\alpha_\sigma$ are expressed as
%
%
\begin{eqnarray}
\alpha_\mu& = &\min \bigl\{1, \exp (\mathrm{SIC}_{\bolds{\phi}^c_\mu, \bolds{\phi}_\sigma}-
\mathrm{SIC}_{\bolds{\phi
}_\mu, \bolds{\phi}_\sigma} )\times R \bigr\}\quad \mbox{and } \label{eqnaccmod2mu}
\\
\alpha_\sigma& =& \min \bigl\{1, \exp (\mathrm{SIC}_{\bolds{\phi}_\mu,\bolds{\phi}^c_\sigma}-
\mathrm{SIC}_{\bolds{\phi}_\mu
,\bolds{\phi}_\sigma} )\times R \bigr\}.\label{eqnaccmod2sigma}
\end{eqnarray}

In order to produce the samples from the posterior distribution of
parameters in $\bolds{\Psi}_a$, we sequentially draw samples for
$\bolds{\phi}_\mu$ and $\bolds{\phi}_\sigma$ by using the two
acceptance probabilities, while marginalizing out ($\bolds{\beta
}$, $\bolds{\theta}$, $\xi$); and then, conditional on the sampled
$\bolds{\phi}_\mu$ and $\bolds{\phi}_\sigma$, draw samples for
($\bolds{\beta}$, $\bolds{\theta}$, $\xi$) using a Normal
approximation based on the maximum likelihood estimates and the
observed information matrix. The detailed simulation procedure can be
found in Step I of Appendix \ref{secIMP}.

\subsection{Submodel 2: Distribution of wind characteristics}\label{subsecenv}

To find a site-specific load distribution, the distribution of wind
characteristics $p(\mathbf{x})$ in (\ref{eqnshort}) needs to be specified.
Since a statistical correlation is noticed between the 10-minute
average wind speed $v$ and the standard deviation of wind speeds $s$ in
Figure~\ref{figSC3}, the distribution of wind characteristics
$p(\mathbf{x})$ can be written as a product of the average wind
speed distribution $p(v)$ and the conditional wind standard deviation
distribution $p(s|v)$. In this section we separately discuss how to
specify each model.

For modeling the 10-minute average wind speed $v$, the IEC standard
suggests using a 2-parameter Weibull distribution (W2) or a Rayleigh
distribution (RAY) [\citet{IEC}]. These two distributions are arguably
the most widely used ones for this purpose. \citet{CRV2008} and \citet
{LS2010} note that under different wind regimes other distributions may
fit wind speed data better, including 3-parameter Weibull
distribution (W3), 3-parameter log-Normal distribution (LN3),
3-parameter Gamma distribution (G3) and 3-parameter inverse-Gaussian
distribution (IG3). We take a total of six candidate distribution
models for average wind speed (W2, W3, RAY, LN3, G3, IG3) from \citet
{LS2010}, and conduct a Bayesian model selection to choose the best
distribution fitting a given average wind speed data set.

We assume UIP priors for the parameters involved in the aforementioned
models, and our approach is again based on maximizing the $\mathrm{SIC}$.
Once the best wind speed model is chosen, we denote it by $\mathcal
{M}_v$. Then, the distribution of 10-minute average wind speed $v$ is
expressed as
%
%
\begin{equation}
v_i \sim \mathcal{M}_v(\bolds{\nu}),\label{eqnv}
\end{equation}
where $\bolds{\nu}$ is the set of parameters specifying
$\mathcal{M}_v$. For instance, if $\mathcal{M}_v$ is W3, then
$\bolds{\nu}=(\nu_1,\nu_2,\nu_3)$, where $\nu_1$, $\nu_2$ and $\nu
_3$ represent the shape, scale and shift parameter, respectively, of a
3-parameter Weibull distribution.

For modeling the standard deviation of wind speed $s$, given the
average wind speed $v$, the IEC standard recommends using a 2-parameter
Truncated Normal distribution (TN2) [\citet{IEC}], which appears to be
what researchers have commonly used; see, for example, \citet{FCV2003}.
The distribution is characterized by a location parameter $\eta$ and a
scale parameter $\delta$. In the literature, both $\eta$ and $\delta$
are treated as a constant. But we observe that data sets measured at
different sites have different relationships between the average wind
speed $v$ and the standard deviation $s$. Some of the $v$-versus-$s$
scatter plots show nonlinear patterns.

Motivated by this observation, we employ a Bayesian MARS model for
modeling $\eta$ and $\delta$, similar to what we did in Submodel 1. The
standard deviation of wind speed $s$, conditional on the average wind
speed $v$, can then be expressed as
%
%
\begin{eqnarray}\label{eqns}
s_i | v_i \sim TN2 \bigl(\eta(v_i),
\delta(v_i) \bigr),
\nonumber
\\[-8pt]
\\[-8pt]
\eqntext{\mbox{where } \eta(v_i)=f_\eta( v_i)
\mbox{ and } \delta (v_i)=\exp \bigl(g_\delta(v_i)
\bigr),}
\end{eqnarray}
where $f_\eta$ and $g_\delta$, like their counterparts in (\ref
{eqnmu1}) and (\ref{eqnsigma1}), are linear combinations of the basis
functions taking the general form (\ref{eqnmu2}). Notice that both of
the functions have only one input variable, which is the average wind speed.

Let $\bolds{\Psi}_{\eta}=(\bolds{\beta}_\eta,\bolds{\phi
}_\eta)$ and $\bolds{\Psi}_{\delta}=(\bolds{\theta}_\delta
,\bolds{\phi}_\delta)$ denote the parameters in $f_\eta(\cdot)$
and $g_\delta(\cdot)$. Since the basis functions $f_\eta$ and $g_\delta
$ in (\ref{eqns}) have only one input variable, only one type of basis
function (i.e., $T_k=1$) is needed. Hence, $\bolds{\phi}_\eta$ and
$\bolds{\phi}_\delta$ are much simpler than $\bolds{\phi}_\mu
$ and $\bolds{\phi}_\sigma$, their counterparts in (\ref
{eqnphimu}) and (\ref{eqnphisigma}), and are expressed as follows:
%
%
\begin{eqnarray}
\label{eqnphieta} \bolds{\phi}_{\eta}= \bigl(K_\eta, \bolds{
\Lambda}^\eta_2, \ldots, \bolds{\Lambda}^\eta_{K_\eta}
\bigr),
\nonumber
\\[-8pt]
\\[-8pt]
\eqntext{\mbox{where } \bolds{\Lambda}^\eta_{k} =
\bigl(T^\eta_{k}, h^\eta _{1k},t^\eta_{1k}
\bigr) \mbox{ and } T^\eta_k=1}
\end{eqnarray}
and
%
%
\begin{eqnarray}\label{eqnphidelta}
\bolds{\phi}_{\delta}= \bigl(K_\delta, \bolds{
\Lambda}^\delta _2, \ldots, \bolds{\Lambda}^\delta_{K_\delta}
\bigr),
\nonumber
\\[-8pt]
\\[-8pt]
\eqntext{\mbox{where } \bolds{\Lambda}^\delta_{k} =
\bigl(T^\delta_{k}, h^\delta_{1k},
t^\delta_{1k} \bigr) \mbox{ and } T^\delta_k=1
.}
\end{eqnarray}

We choose the prior distribution for $(\bolds{\beta}_\eta,
\bolds{\theta}_\delta)$ as UIP and the prior for $(\bolds{\phi
}_\eta, \bolds{\phi}_\delta)$ as uniform distribution, and solve
this Bayesian MARS model by using a RJS algorithm, as in the preceding
two sections. The predictive distributions of the average wind speed
$\tilde{v}$ and the standard deviation $\tilde s$ are
%
%
\begin{eqnarray}
p(\tilde v|\mathcal{D}_v) &=& \int p(\tilde v| \bolds{\nu}) p(
\bolds{\nu}| \mathcal{D}_v) \,d \bolds{\nu }\quad \mbox{and}
\label{eqnvdist}
\\
\qquad p(\tilde s|\tilde v,\mathcal{D}_v,\mathcal{D}_s) &=&
\int\int p(\tilde s|\tilde v,\bolds{\Psi}_{\eta},\bolds{
\Psi}_{\delta
})p(\bolds{\Psi}_{\eta},\bolds{\Psi}_{\delta}|
\mathcal {D}_v,\mathcal{D}_s ) \,d\bolds{
\Psi}_{\eta}\,d\bolds{\Psi }_{\delta}, \label{eqnsdist}
\end{eqnarray}
where $\mathcal{D}_v$ and $\mathcal{D}_s$ are the data sets of the
observed average wind speeds and the standard deviations. The detailed
simulation procedure is included in Step II in Appendix \ref{secIMP}.

\subsection{Posterior predictive distribution of the extreme load level
$l_T$}\label{subsecextreme}
We are interested in getting the posterior predictive distribution of
the quantile value $l_T$, based on the observed load and wind data
$\mathcal{D}:=(\mathcal{D}_y, \mathcal{D}_{v},\mathcal{D}_{s})$. In
order to do so, we need to draw samples $\tilde{y}$'s from the
predictive distribution of the maximum load given parameters $p
[\tilde{y}|\mathcal{D}, \bolds{\Psi}_a  ]$, which is
%
%
\begin{equation}
p [\tilde{y}|\mathcal{D},\bolds{\Psi}_a ] = \int\int p[\tilde{y}|
\tilde{v},\tilde{s},\bolds{\Psi}_a,\mathcal {D}]p[\tilde{v},
\tilde{s}|\mathcal{D}_v,\mathcal{D}_s] \,d\tilde{v}\,d
\tilde{s},
\end{equation}
where $p[\tilde{v},\tilde{s}|\mathcal{D}_v,\mathcal{D}_s]$ can be
expressed as the product of (\ref{eqnvdist}) and
(\ref{eqnsdist}).\vadjust{\goodbreak}

To calculate a quantile value of the load for a given $P_T$ [as in (\ref
{eqnpt})], we go through the following steps:
\begin{itemize}
\item Draw samples from the joint posterior predictive distribution
$p [\tilde{v},\tilde{s}|\mathcal{D}_{v},\mathcal{D}_{s} ]$ of
wind characteristics $(\tilde{v},\tilde{s})$ (Step II in Appendix \ref
{secIMP});
\item Draw a set of samples from the posterior distribution of model
parameters $\bolds{\Psi}_a = (\bolds{\Psi}_{\mu},\bolds{\Psi}_{\sigma},\xi)$;
this is realized by employing the RJS algorithm
in Section~\ref{subsecposterior} (or Step I in Appendix \ref{secIMP});
\item Given the above samples of wind characteristics and model
parameters, we calculate ($\mu, \sigma, \xi$) that are needed in a GEV
distribution; this yields a short-term distribution $p [\tilde
{y}|\tilde{v},\tilde{s},\bolds{\Psi}_a ]$;
\item Integrating out the wind characteristics $(\tilde{v},\tilde{s})$,
obtain the long-term distribution $p [\tilde{y}|\mathcal
{D},\bolds{\Psi}_a ]$;
\item Draw samples from $p [\tilde{y}|\mathcal{D},\bolds{\Psi
}_a ]$, and compute a quantile value $l_T[\bolds{\Psi}_a]$
corresponding to $P_T$.
\end{itemize}

In fact, the predictive mean and Bayesian credible interval of the
extreme load level $l_T$ are obtained when running the RJS algorithm.
The RJS runs through $M_l$ iterations and, at each iteration, we obtain
a set of samples of the model parameters $\bolds{\Psi}_a$ and
calculate a $l_T[\bolds{\Psi}_a]$. Once $M_l$ values of
$l_T[\bolds{\Psi}_a]$ are obtained, its mean and credible
intervals can then be numerically computed.

\begin{table}[b]
\caption{$\mathrm{SIC}$ for the average wind speed
models}\label{tblresult2}
\begin{tabular*}{250pt}{@{\extracolsep{\fill}}lccc@{}}
\hline
\textbf{Distributions} & \textbf{ILT1} & \textbf{ILT2} &
\multicolumn{1}{c@{}}{\textbf{ILT3}}\\
\hline
W2 & $-2984$ & $-1667$ & $-12\mbox{,}287$\\
W3 & $\mathbf{-2941}$ & $\mathbf{-1663}$ & $\mathbf{-11\mbox{\textbf{,}}242}$
\\
RAY & $-3120$ & $-1779$ & $-13\mbox{,}396$\\
LN3 & $-2989$ & $-1666$ & $-11\mbox{,}444$\\
G3 & $-2974$ & $-1666$ & $-11\mbox{,}290$\\
IG3 & $-2986$ & $-2313$ & $-11\mbox{,}410$\\
\hline
\end{tabular*}
\end{table}

\section{Results}\label{secresult}

\subsection{Model selection}\label{subsecmodres}
Table~\ref{tblresult2} presents the $\mathrm{SIC}$ values of the six
candidate average wind speed models using different ILT data sets. The
boldfaced values indicate the largest $\mathrm{SIC}$ for a given data set
and, consequently, the corresponding models are chosen for that data set.

Regarding the average wind speed model, all candidate distributions
except RAY provide generally a good model fit for ILT1, with a similar
level of fitting quality, but W3 dominates slightly. For the ILT2 data,
W2, W3, LN3 and G3 produce similar $\mathrm{SIC}$ values. In the ILT3
data, W3, LN3, G3 and IG3 perform similarly. Still W3 is slightly
better. So we choose W3 as our average wind speed model.

\begin{figure}

\includegraphics{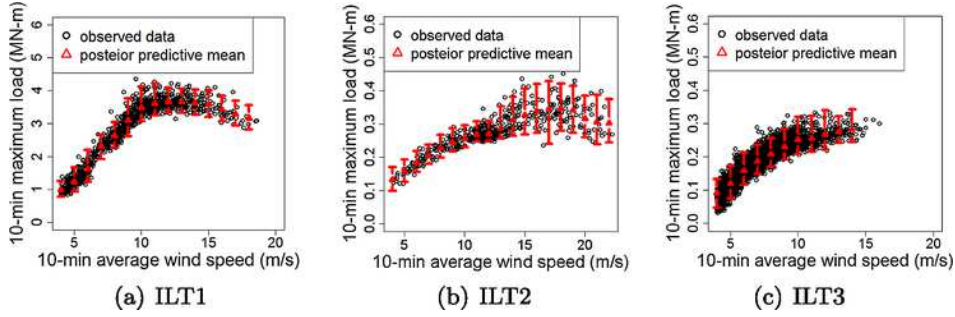}

\caption{95\% point-wise credible intervals for different wind speeds.} \label{figCI1}
\end{figure}

\begin{figure}[b]

\includegraphics{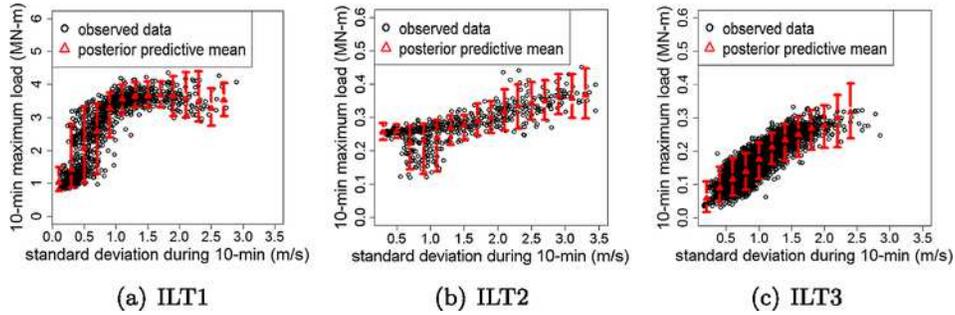}

\caption{95\% point-wise credible intervals for
different standard deviations.} \label{figCI2}
\end{figure}

\subsection{Point-wise credible intervals}\label{subsecPCI}

As a form of checking the conditional maximum load model, we present in
Figures~\ref{figCI1} and \ref{figCI2} the 95\% point-wise credible
intervals under different wind speeds and standard deviations. To
generate these figures, we take a data set and fix $v$ or $s$ at one
specific speed or standard deviation at a time and then draw the
posterior samples for $\tilde{y}$ from the posterior predictive
distribution of conditional maximum load, $p(\tilde{y}|\mathbf{x})$. Suppose that we want to generate the credible intervals at wind
speed $v_*$ or standard deviation $s_*$. The posterior predictive
distributions are computed as follows:
\begin{eqnarray}
p \bigl(\tilde{y}|(v,s)\in\mathcal{D}_{v_*},\mathcal{D}_y
\bigr)&=&\int p \bigl(\tilde {y}|(v,s)\in\mathcal{D}_{v_*}, \bolds{
\Psi}_a \bigr) p(\bolds{\Psi }_a|\mathcal{D}_y)
\,d\bolds{\Psi}_a,
\nonumber
\label{eqnCI1}
\\
p \bigl(\tilde{y}|(v,s)\in\mathcal{D}_{s_*},\mathcal{D}_y
\bigr)&=&\int p \bigl(\tilde {y}|(v,s)\in\mathcal{D}_{s_*}, \bolds{
\Psi}_a \bigr) p(\bolds{\Psi }_a|\mathcal{D}_y)
\,d\bolds{\Psi}_a,
\nonumber
\label{eqnCI2}
\end{eqnarray}
where $\mathcal{D}_{v_*}$ and $\mathcal{D}_{s_*}$ are subsets
of the observed data such that $\mathcal{D}_{v_*} =\{(v_i,s_i) \dvtx\break v_*-0.5 < v_i < v_*+0.5, \mbox{ and } (v_i,s_i) \in\mathcal{D}_{v,s}\}
$ and $\mathcal{D}_{s_*} =\{(v_i,s_i) \dvtx  s_*-0.05 < s_i < s_*+0.05, \mbox{ and } (v_i,s_i) \in\mathcal{D}_{v,s}\}$. Given these distributions,
samples for $\tilde{y}$ are drawn to construct the $95\%$ credible
intervals at $v_*$ or $s_*$. The result is shown as one vertical bar in
either a $v$-plot (Figure~\ref{figCI1}) or a $s$-plot (Figure~\ref{figCI2}). To complete these figures, the process is repeated in the
$v$-domain with 1 m$/$s increment and in the $s$-domain with 0.2 m$/$s
increment. These figures show that the variability in data are
reasonably captured by the spline method.

\subsection{Comparison between the binning method and spline method for
conditional maximum load}\label{subseccom}

In our procedure for estimating the extreme load level, two different
distributions of maximum load $y$ are involved: one is the conditional
maximum load distribution $p(y|\mathbf{x})$, aka the short-term
distribution, and the other is the unconditional maximum load
distribution $p(y)$, aka the long-term distribution. Using the observed
field data, it is difficult to assess the estimation accuracy of the
extreme load levels in the long-term distribution, because of the
relatively small amount of observation records. What we undertake in
this section is to evaluate a method's performance of estimating the
tail of the short-term distribution $p(y|\mathbf{x})$. We argued
before that the short-term distribution underlies the difference
between the proposed Bayesian spline method and the binning method. The
comparison in this section is intended to show the advantage of the
Bayesian spline method. In Section~\ref{subsecsimulation} we employ a
simulation study that generates a much larger data set, allowing us to
compare the performance of two methods in estimating the extreme load
level in the long-term distribution.

To evaluate the tail part of a conditional maximum load distribution,
we compute a set of upper quantile estimators and assess their
estimation qualities using the generalized piecewise linear (GPL) loss
function [\citet{Gneiting2011a}]. A GPL is defined as follows:
%
\begin{eqnarray}\qquad
\label{eqngpl1} &&S_{\tau, b} \bigl(\hat{l}(\mathbf{x}_i),y(
\mathbf{x}_i) \bigr)
\nonumber
\\[-8pt]
\\[-8pt]
\nonumber
&&\qquad=\cases{\displaystyle %
\bigl(\mathbh{1} { \bigl(\hat{l}(
\mathbf{x}_i)\geq y(\mathbf{x}_i) \bigr)}-\tau \bigr)
\frac{1} {|b|} \bigl( \bigl[\hat{l}(\mathbf{x}_i)
\bigr]^b- \bigl[y(\mathbf{x}_i) \bigr]^b
\bigr),&\quad $\mbox{for } b \neq{0}$,
\vspace*{2pt}\cr
\displaystyle\bigl(\mathbh{1} { \bigl(\hat{l}(\mathbf{x}_i) \geq y(
\mathbf{x}_i) \bigr)}-\tau \bigr) \log{ \biggl(\frac{\hat{l}(\mathbf{x}_i)} {y(\mathbf{x}_i)}
\biggr)},&\quad $\mbox{for } b = {0},$}
\end{eqnarray}
where $\hat{l}(\mathbf{x}_i)$ is the $\tau$-quantile
estimation of $p(y|\mathbf{x}_i$) for a given $\mathbf{x}_i$,
$y(\mathbf{x}_i)$ is the observed maximum load in the test data
set, given the same $\mathbf{x}_i$, $b$ is a power parameter, and
$\mathbh{1}$ is an indicator function. The power parameter $b$ usually
ranges between 0 and 2.5. When $b=1$, the GPL loss function is the same
as the piecewise linear (PL) loss function.

For the above empirical evaluation, we randomly divide a data set into
a partition of 80\% for training and 20\% for testing. We use the
training set to establish a short term distribution $p(y|\mathbf{x}$). For any $\mathbf{x}_i$ in the test set, the $\tau$-quantile
estimation $\hat{l}(\mathbf{x}_i)$ can be computed using
$p(y|\mathbf{x}$). And then, the GPL loss function value is taken
as the average of all $S_{\tau, b}$ values over the test set, as follows:
%
%
\begin{equation}
\overline{S}_{\tau, b}=\frac{1}{n_t}\sum_{i=1}^{n_{t}}S_{\tau, b}
\bigl(\hat {l}_i(\mathbf{x}_i),y_i \bigr),
\label{eqngpl2}
\end{equation}
where $n_{t}$ is the number of data points in a test set and
$y_i$ is the same as $y(\mathbf{x}_i)$. We call $\overline{S}_{\tau
, b}$ the mean score. We repeat the training/test procedure 10 times,
and the final mean score is the average of the ten mean scores. For
notational simplicity, we still call the final mean score the mean
score and use $\overline{S}_{\tau, b}$ to represent it, as long as its
meaning is clear in the context.

In this comparison, we use two methods to establish the short-term
distribution: the binning method and the proposed Bayesian spline
method. In our RJS algorithm in Section~\ref{subsecposterior}, we draw
$N_l=100$ samples from the short-term distribution. Accordingly, we can
evaluate the quality of quantile estimations of the short-term
distribution for a $\tau$ up to $0.99$.

\begin{figure}[b]

\includegraphics{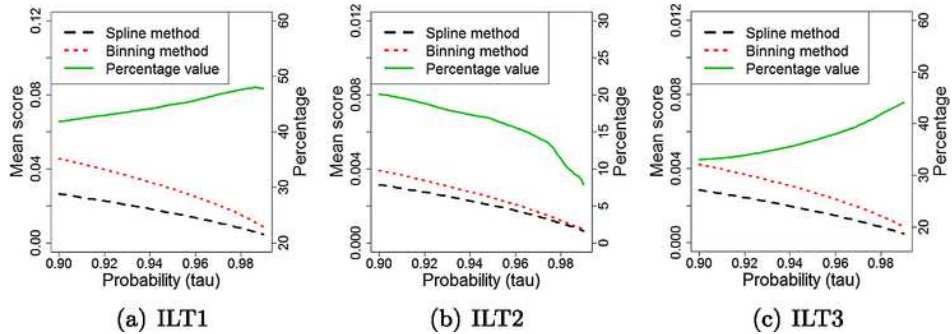}

\caption{Comparison of PL function: the left Y-axis
represents the mean score values and the right Y-axis represents the
percentage values, which are the reduction in the mean scores when the
spline method is compared with the binning method.} \label{figgpl}
\end{figure}

We first take a look at the comparisons in Figure~\ref{figgpl}, which
compares the PL loss (i.e., $b=1$) of both methods as $\tau$ varies in
the above-mentioned range. The left vertical axis shows the values of
the mean score of the PL loss, while the right axis is the percentage
value of the reduction in mean scores when the spline method is
compared with the binning method. For all three data sets, the spline
method maintains lower mean scores than the binning method.

When $\tau$ is approaching $0.99$ in Figure~\ref{figgpl}, it looks
like the PL losses of the spline and binning methods are getting closer
to each other. This is largely due to the fact that the PL loss values
are smaller at a higher $\tau$, so that their differences are
compressed in the figure. If one looks at the solid line in a plot,
which represents the percentage of reduction in the mean score, the
spline method's advantage over the binning method is more evident in
the cases of ILT1 and ILT3 data sets. When $\tau$ gets larger, the
spline method produces a significant improvement over the binning
method, with a reduction of PL loss ranging from $33\%$ to $50\%$. The
trend is different when using the ILT2 data set. But still, the spline
method can reduce the mean scores of the PL loss from the binning
method by $8\%$ to $20\%$. Please note that the ILT2 data set is the
smallest set, having slightly fewer than 600 data records. We believe
that the difference observed over the ILT2 case is attributable to the
scarcity of data.

We compute the mean scores of the GPL loss under three different power
parameters $b=0,1,2$ for each method. Table~\ref{tblresult3} presents
the results under $\tau=0.9$, while Table~\ref{tblresult4} is for $\tau
=0.99$. In Table~\ref{tblresult3} the spline method has a mean score
20\% to 42\% lower than the binning method. In Table~\ref{tblresult4}
the reductions in mean scores are in a similar range. Overall, these
results clearly show the improvement achieved by employing the Bayesian
spline method.

\begin{table}
\caption{Mean scores of GPL/PL for the $0.9$-quantile
estimators}\label{tblresult3}
\begin{tabular*}{\textwidth}{@{\extracolsep{\fill}}lcccccc@{}}
\hline
& \multicolumn{2}{c}{\textbf{ILT1}} & \multicolumn{2}{c} {\textbf{ILT2}} & \multicolumn
{2}{c@{}} {\textbf{ILT3}}\\[-6pt]
& \multicolumn{2}{c}{\hrulefill} & \multicolumn{2}{c} {\hrulefill} & \multicolumn
{2}{c@{}} {\hrulefill}\\
{\textbf{Power parameter}} & \textbf{Binning} & \textbf{Spline} & \textbf{Binning}& \textbf{Spline}& \textbf{Binning}&
\textbf{Spline}\\
\hline
$b=0$ & 0.0185 & 0.0108 & 0.0129 & 0.0103& 0.0256 & 0.0171
\\
$b=1$ & 0.0455 & 0.0265 & 0.0040 & 0.0031 & 0.0042 &
0.0028\\
$b=2$ & 0.1318& 0.0782 & 0.0013 & 0.0010 &
0.0008& 0.0005 \\
\hline
\end{tabular*}
\end{table}

\begin{table}[b]
\caption{Mean scores of GPL/PL for the $0.99$-quantile
estimators}\label{tblresult4}
\begin{tabular*}{\textwidth}{@{\extracolsep{\fill}}lcccccc@{}}
\hline
& \multicolumn{2}{c}{\textbf{ILT1}} & \multicolumn{2}{c} {\textbf{ILT2}} & \multicolumn
{2}{c@{}} {\textbf{ILT3}}\\[-6pt]
& \multicolumn{2}{c}{\hrulefill} & \multicolumn{2}{c} {\hrulefill} & \multicolumn
{2}{c@{}} {\hrulefill}\\
{\textbf{Power parameter}} & \textbf{Binning} & \textbf{Spline} & \textbf{Binning}& \textbf{Spline}& \textbf{Binning}&
\textbf{Spline}\\
\hline
$b=0$ & 0.0031 & 0.0018 & 0.0022 & 0.0020 & 0.0045 & 0.0027
\\
$b=1$ & 0.0086 & 0.0045 & 0.0007 & 0.0006 & 0.0008 &
0.0005\\
$b=2$ & 0.0270 & 0.0135 & 0.0003 & 0.0002 &
0.0002 & 0.0001 \\
\hline
\end{tabular*}
\end{table}

\begin{figure}

\includegraphics{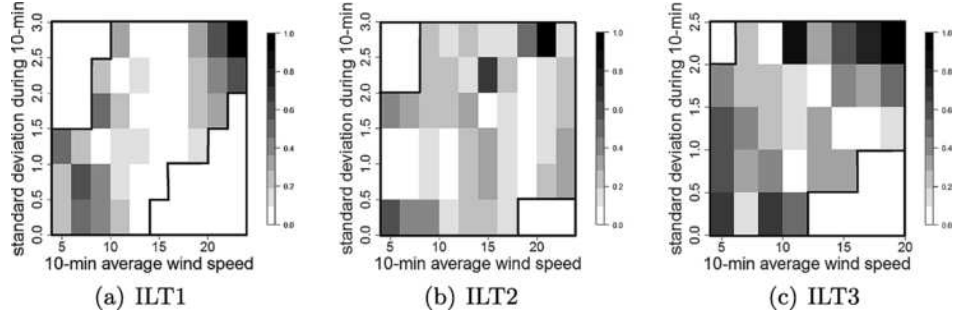}

\caption{Comparison of the $0.99$-quantiles between binning
method and spline method.} \label{figIMG}
\end{figure}

In order to understand the difference between the spline method and
binning method, we compare the $0.99$ quantiles of the 10-minute
maximum load conditional on a specific wind condition. This is done by
computing the difference in the quantile values of the conditional
maximum load from the two methods for different weather bins. The wind
condition of each bin is approximated by the median values of $v$ and
$s$ in that bin. Figure~\ref{figIMG} shows the standardized difference
of the two $0.99$ quantile values in each bin. The darker the color is,
the bigger the difference. Note that we exclude comparisons in the
weather bins with very low likelihood, namely, low wind speed and high
standard deviation or high wind speed and low standard deviation.

%
\begin{table}[b]
\caption{Estimates of extreme load levels ($l_{T}, T=20$
years), unit: MN-m}\label{tblresult6}
\begin{tabular*}{\textwidth}{@{\extracolsep{\fill}}lcc@{}}
\hline
\textbf{Data sets} & \textbf{Binning method} & \textbf{Spline method}\\
\hline
ILT1 & 6.455 (6.063, 7.092) & 4.750 (4.579, 4.955)\\
ILT2 & 0.752 (0.658, 0.903) & 0.576 (0.538, 0.627)\\
ILT3 & 0.505 (0.465, 0.584) & 0.428 (0.398, 0.463)\\
\hline
\end{tabular*}
\end{table}

\begin{table}[b]
\caption{Estimates of extreme load level ($l_{T}, T=50$
years), unit: MN-m}\label{tblresult7}
\begin{tabular*}{\textwidth}{@{\extracolsep{\fill}}lcc@{}}
\hline
\textbf{Data sets} & \textbf{Binning method} & \textbf{Spline method}\\
\hline
ILT1 & 6.711 (6.240, 7.485) & 4.800 (4.611, 5.019)\\
ILT2 & 0.786 (0.682, 0.957) & 0.589 (0.547, 0.646)\\
ILT3 & 0.527 (0.480, 0.621) & 0.438 (0.405, 0.476)\\
\hline
\end{tabular*}
\end{table}

We can observe that the two methods produce similar results at the bins
having a sufficient number of data points (mostly weather bins in the
central area), and the results are different when the data are scarce---this
tends to happen at the two ends of the average wind speed and
standard deviation. This echoes the point we made earlier that without
binning the weather conditions, the spline method is able to make
better use of the available data and overcome the limited data problem
for rare weather events.

\subsection{Estimation of extreme load}\label{subsecextremeres}
Finally, Tables~\ref{tblresult6} and \ref{tblresult7} show the
estimates of the extreme load levels $l_T$, corresponding to $T = 20$
and $T = 50$ years, respectively. The values in parenthesis are the
95\% credible (or confidence) intervals.
We observe that the extreme load levels $l_T$ obtained by the binning
method are generally higher than those obtained by the spline method.
This should not come as a surprise. As we push for a high quantile,
more data would be needed in each weather bin, but the amounts in
reality are limited due to the binning method's compartmentalization of
data. The binning method also produces a wider confidence interval than
the spline method, as a result of the same rigidity in data handling.
The detailed procedure for computing the binning method's confidence
interval is included in Appendix \ref{secPROC}.

\subsection{Simulation of extreme load}\label{subsecsimulation}

In this section a simulation study is undertaken to assess the
estimation accuracy of extreme load level in the long-term
distribution. The simulations use one single covariate $x$, mimicking
the wind speed, and a dependent variable $y$, corresponding to the
maximum load. We use the following procedure to generate the simulated data:

\begin{longlist}[(a)]
\item[(a)] Generate a sample $x_{i}$ from a 3-parameter Weibull
distribution. Then sample $x_{ij}$, $j =1,\ldots,1000$, from a normal
distribution having $x_{i}$ as its mean and a unit variance. The set of
$x_{ij}$'s represents the different wind speeds within a bin.
\item[(b)] Draw the samples $y_{ij}$ from a normal distribution with its
mean as $\mu^s_{ij}$ and its standard deviation as $\sigma^s_{ij}$,
which are expressed as follows:
\begin{eqnarray*}
\mu^s_{ij} &=& \cases{ %
\displaystyle\frac{1.5} {[1+48 \times\exp(-0.3 \times x_{ij})]},\vspace*{2pt}\cr
\qquad \mbox{if } x_{i} < 17,
\vspace*{2pt}\cr
\displaystyle\frac{1.5} {[1+48 \times\exp(-0.3 \times x_{ij})]} + \bigl[0.5-0.0016 \times \bigl(x_{i}+x_{i}^2
\bigr) \bigr],\vspace*{2pt}\cr
\qquad\mbox{if } x_{i} \geq17,}\\
\sigma^s_{ij} &=& 0.1 \times\log(x_{ij}).
\end{eqnarray*}
The above set of equations is used to create a $y$ response resembling
the load data we observe. The parameters used in the equations are
chosen through trials so that the simulated $y$ looks like the actual
mechanical load response. While many of the parameters used above do
not have any physical meaning, some of them do, for instance, the
``$17$'' in ``$x_{i} < 17$'' bears the meaning of the rated wind speed.
\item[(c)] Find the maximum value $y_i=\max\{y_{i,1},\ldots,y_{i,1000}\}$,
corresponding to $x_i$. According to the classical extreme value theory
[\citet{Coles2001,Smith1990}], $y_i$ produced in such a way can be
modeled by a GEV distribution.
\item[(d)] Repeat (a) through (c) for $i=1, \ldots, 1000$ to produce the
training data set with $n=1000$ data pairs, and denote this data set
by $\mathcal{D}_{\mathrm{TR}}=\{(x_1,y_1),\ldots, (x_{1000},y_{1000})\}$.
\end{longlist}

Once the training data set $\mathcal{D}_{\mathrm{TR}}$ is simulated, both the
binning method and spline method are used to estimate the extreme load
levels $l_T$ corresponding to two probabilities: $0.0001$ and
$0.00001$. This estimation is based on drawing samples from the
long-term distribution of $y$, as described in Section~\ref{subsecextreme}, which produces the posterior predictive distribution
of $l_T$. To compare the estimation accuracy of the extreme quantile
values, we also generate 100 simulated data sets; each data set
consists of $100\mbox{,}000$ data points, which are obtained by repeating the
above (a) through (c). For each data set, we find the observed quantile
values $l_{0.0001}$ and $l_{0.00001}$. Using the 100 simulated data
sets, we also obtain 100 different samples of these quantiles.

\begin{figure}

\includegraphics{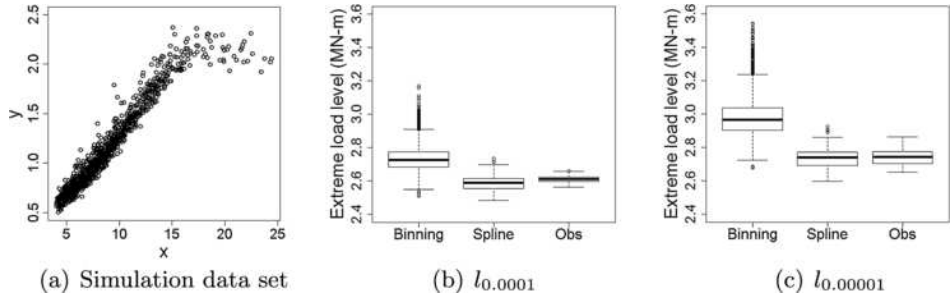}

\caption{Simulation data set, estimated and observed extreme
quantile values: \textup{(a)} An example of a simulated data set. \textup{(b)} and \textup{(c)}
Boxplots for the distribution of the binning estimate, the Bayesian
spline estimate and the respective sample quantile across 100 simulated
data sets.} \label{figSIM}
\end{figure}

Figure~\ref{figSIM}(a) shows a scatter plot of the simulated $x$'s and
$y$'s in $\mathcal{D}_{\mathrm{TR}}$, which resembles the load responses we saw
previously. Figure~\ref{figSIM}(b) and (c) present the extreme load
levels estimated by the two methods as well as the observed extreme
quantile values under the two selected probabilities. We observe that
the binning method tends to overestimate the extreme quantile values
and yields wider confidence intervals than the spline method.
Furthermore, the degree of overestimation appears to increase as the
probability corresponding to an extreme quantile value goes smaller.
This observation confirms what we observed in Section~\ref{subsecextremeres} using the field data. This simulation result
suggests that using the binning method for extreme load estimation is
not a good practice.
%

\section{Summary}\label{secsummary}
This study presents a Bayesian spline method for estimating the extreme
load on wind turbines. The spline method essentially supports a
nonhomogeneous GEV distribution to capture the nonlinear relationship
between the load response and the wind-related covariates. Such
treatment avoids binning the data. The underlying spline models instead
connect all the bins across the whole wind profile, so that load and
wind data are pooled together to produce better estimates. This is
demonstrated by applying the spline method to three sets of inland wind
turbine load response data and making comparisons with the binning method.

The popularity of the binning method in industrial practice is due to
the simplicity of its idea and procedure. However, simplicity of a
procedure should not be mistaken as simplicity of a model. 
Suppose that one uses a $6 \times10$ grid to bin the two-dimensional
wind covariates (as we did in this study) and fixes the shape parameter
$\xi$ across the bins (a common practice in the industry). The binning
method yields $60$ local GEV distributions, each of which has two
parameters, translating to a total of 121 parameters for the overall
model (counting the fixed $\xi$ as well). By contrast, the spline
method, although conceptually and procedurally more involved, produces
an overall model with fewer parameters. To see this, consider the
following: for the three ILT data sets, the average ($K_{\mu} +
K_{\sigma}$) from the RJS algorithm is between $12$ and $18$. The
number of model parameters $d_k$ in (\ref{eqnSIC}) is generally less
than $20$, a number far smaller than the number of parameters in the
binning method. In the end, the spline method uses a sophisticated
procedure to find a simpler model that is more capable.

\begin{appendix}
\section{Priors}\label{secPRI}
In this appendix we specify priors for parameters used in the basis
functions as follows:
\begin{eqnarray}
\bolds{\phi}&=& (K, \bolds{\Lambda}_2, \ldots, \bolds{
\Lambda}_{K} )
\nonumber
\\[-8pt]
\\[-8pt]
\eqntext{\mbox{where } \bolds{\Lambda}_{k} = \cases{ %
 (T_{k}, h_{1k}, t_{1k}),&\quad
$\mbox{when } T_k=1, 2$,
\vspace*{2pt}\cr
(T_{k}, h_{1k}, h_{2k},t_{1k},
t_{2k}),&\quad $\mbox{when } T_k=3,$}}\\
%
p(K)&=&\frac{1}{n}, \qquad K= \{1,\ldots, n \},
\nonumber
\\
p(T_{k})&=& \cases{ %
1,&\quad $T_{k}= \{1 \},\mbox{for } \bolds{ \phi}_{\eta} \mbox{ and
} \bolds{\phi }_{\delta}$,
\vspace*{2pt}\cr
\frac{1}{2}, &\quad $T_{k}= \{1,2 \},\mbox{for } \bolds{
\phi}_{\mu} \mbox{ in ILT2 and all } \bolds{ \phi}_{\sigma}$,
\vspace*{2pt}\cr
\frac{1}{3},&\quad $ T_{k}= \{1, 2, 3 \},\mbox{for } \bolds{
\phi}_{\mu} \mbox{ in ILT1 and ILT3},$}
\nonumber\\
p(h_{\cdot k}) &=& \tfrac{1}{2},\qquad h_{\cdot k}= \{+1,-1 \},
\nonumber
\\
p(t_{\cdot k}) &=& \frac{1} {n}, \qquad t_{\cdot k}= \{v_1,
\ldots, v_n \} \mbox{ or } \{s_1, \ldots,
s_n \}.
\nonumber
\end{eqnarray}

\section{Implementation details of the spline method}\label{secIMP}
In this appendix we provide the detailed implementation procedure for
the spline method. The procedure consists of two major steps: (1) Step
I: construct the posterior predictive distribution of the extreme load
level $l_T$ and (2) Step II: obtain the joint posterior predictive
distribution of wind characteristics $(v,s)$.

\begin{enumerate}
\item\textit{Step} I: construct the posterior predictive distribution
of the extreme load level using the Bayesian spline models:

\begin{enumerate}[(a)]
\item[(a)] Set $t=0$ and the initial $\bolds{\phi}^{(t)}_{\mu}$ and
$\bolds{\phi}^{(t)}_{\sigma}$ both to be a constant scalar.
\item[(b)] At iteration $t$, $K_\mu$ and $K_\sigma$ are equal to the number
of basis functions specified in $\bolds{\phi}^{(t)}_{\mu}$ and
$\bolds{\phi}^{(t)}_{\sigma}$. Find the MLEs of $\bolds{\beta
}^{(t)}, \bolds{\theta}^{(t)}, \xi^{(t)}$ and the inverse of the
negative of Hessian matrix, given $\bolds{\phi}^{(t)}_{\mu}$ and
$\bolds{\phi}^{(t)}_{\sigma}$.
\item[(c)] Generate $u^1_\mu$ uniformly on $[0,1]$ and choose a move in the
RJS procedure. In the following, $b_{K_\mu}, r_{K_\mu}, m_{K_\mu}$ are
the proposal probabilities associated with a move type, and they are
all set as $\frac{1}{3}$:
\begin{itemize}
\item If $(u^1_\mu\leq b_{K_\mu})$, then go to \rmfamily{BIRTH} step,
denoted by $\bolds{\phi}_\mu^{*}=$ BIRTH-proposal$(\bolds{\phi
}_\mu^{(t)})$, which is to augment $\bolds{\phi}_\mu^{(t)}$ with a
$\bolds{\Lambda}^\mu_{K_\mu+1}$ that is selected uniformly at random;
%
\item Else if $(b_{K_\mu}\leq u^1_\mu\leq b_{K_\mu}+r_{K_\mu})$,
then go to \rmfamily{DEATH} step, denoted by $\bolds{\phi}_\mu
^{*}=$ DEATH-proposal$(\bolds{\phi}_\mu^{(t)})$, which is to
remove from $\bolds{\phi}_\mu^{(t)}$ with a $\bolds{\Lambda}
^\mu_{k}$ where $2\leq k \leq K_\mu$ is selected uniformly at random;
%
\item Else, go to \rmfamily{MOVE} step, denoted by $\bolds{\phi}_\mu
^{*}=$MOVE-proposal$(\bolds{\phi}_\mu^{(t)})$, which first does
$\bolds{\phi}_\mu^{\dag}$ = DEATH-proposal$(\bolds{\phi}_\mu
^{(t)})$ and then does $\bolds{\phi}_\mu^{*}$ =
BIRTH-proposal$(\bolds{\phi}_\mu^{\dag})$.
\end{itemize}
\item[(d)] Find the MLEs ($\bolds{\beta}^{*}, \bolds{\theta}^{*},
\xi^{*}$) and the inverse of the negative of Hessian matrix, given
$\bolds{\phi}^{*}_{\mu}$ and $\bolds{\phi}_{\sigma}$.
\item[(e)] Generate $u^2_\mu$ uniformly on $[0,1]$ and compute the
acceptance ratio $\alpha_{\mu}$ in~(\ref{eqnaccmod2mu}), using the
results from (b) and (d).
\item[(f)] Accept $\bolds{\phi}_\mu^{*}$ as $\bolds{\phi}_\mu
^{(t+1)}$ with probability $\min(\alpha_{\mu},1)$. If $\bolds{\phi
}_\mu^{*}$ is not accepted, let $\bolds{\phi}_\mu
^{(t+1)}=\bolds{\phi}_\mu^{(t)}$.

\item[(g)] Generate $u_\sigma^1$ uniformly on $[0,1]$ and choose a move in
the RJS procedure. In the following, $b_{K_\sigma}, r_{K_\sigma},
m_{K_\sigma}$ are the proposal probabilities associated with a move
type, and they are all set as $\frac{1}{3}$:
\begin{itemize}
\item If $(u_\sigma^1\leq b_{K_\sigma})$, then go to \rmfamily{BIRTH}
step, denoted by $\bolds{\phi}_\sigma^{*}=$
BIRTH-proposal$(\bolds{\phi}_\sigma^{(t)})$, which is to augment
$\bolds{\phi}_\sigma^{(t)}$ with a $\bolds{\Lambda}^\sigma
_{K_\sigma+1} $ that is selected uniformly at random;
%
\item Else if $(b_{K_\sigma}\leq u^1_\sigma\leq b_{K_\sigma
}+r_{K_\sigma})$, then go to \rmfamily{DEATH} step, denoted by
$\bolds{\phi}_\sigma^{*}=$ DEATH-proposal$(\bolds{\phi}_\sigma
^{(t)})$, which is to remove from $\bolds{\phi}^{(t)}$ with a
$\bolds{\Lambda}^\sigma_{k} $ where $2\leq k \leq K_\sigma$ that is
selected uniformly at random;
%
\item Else, go to \rmfamily{MOVE} step, denoted by $\bolds{\phi
}_\sigma^{*}=$MOVE-proposal$(\bolds{\phi}_\sigma^{(t)})$, which
first does $\bolds{\phi}_\sigma^{\dag}$ = DEATH-proposal$(\bolds{\phi}^{(t)})$ and then does $\bolds{\phi}_\sigma^{*}$ =
BIRTH-proposal$(\bolds{\phi}_\sigma^{\dag})$.
\end{itemize}

\item[(h)] Find the MLEs ($\bolds{\beta}^{*}, \bolds{\theta}^{*},
\xi^{*}$) and the inverse of the negative of Hessian matrix, given
$\bolds{\phi}^{t+1}_{\mu}$ and $\bolds{\phi}^{*}_{\sigma}$.
\item[(i)] Generate $u_\sigma^2$ uniformly on $[0,1]$ and compute the
acceptance ratio $\alpha_{\sigma}$ in~(\ref{eqnaccmod2sigma}), using
the results from (d) and (h).
\item[(j)] Accept $\bolds{\phi}_\sigma^{*}$ as $\bolds{\phi}_\sigma
^{(t+1)}$ with probability $\min(\alpha_{\sigma},1)$. If $\bolds{\phi}_\sigma^{*}$ is not accepted, let $\bolds{\phi}_\sigma
^{(t+1)}=\bolds{\phi}_\sigma^{(t)}$.

\item[(k)] After initial burn-ins (in our implementation, initial burn-in is
1000), draw a posterior sample of $(\bolds{\beta}^{(t+1)},
\bolds{\theta}^{(t+1)}, \xi^{(t+1)})$ from the approximated
multivariate normal distribution at the maximum likelihood estimates
and the inverse of the negative of the Hessian matrix. Depending on the
acceptance or rejection that happened in (f) and (j), the MLEs to be
used are obtained from either (b), (d) or (h).

\item[(l)] Take the posterior sample of $\bolds{\Psi}_a$, obtained in
(f), (j) and (k), and calculate a sample of $\mu$ and $\sigma$ using
(\ref{eqnmu1}) and (\ref{eqnsigma1}), respectively, for each pair of
the $N_w \times N_{sw}$ samples of ($v, s$) obtained in Step II. This
generates $N_w \times N_{sw}$ samples of $\mu$ and $\sigma$.

\item[(m)] Draw $N_l$ samples for the 10-minute maximum load $\tilde{y}$
from each GEV distribution with $\mu_i$, $\sigma_i$ and $\xi_i$,
$i=1,\ldots, N_w \times N_{sw}$, where $\mu_i$ and $\sigma_i$ are among
$N_w \times N_{sw}$ samples obtained in (l), and $\xi_i$ is always set
as $\xi^{(t+1)}$.
\item[(n)] Get the quantile value (i.e., the extreme load level
$l_T[\bolds{\Psi}_a]$) corresponding to $1-P_T$ from the $N_w
\times N_{sw} \times N_l$ samples of $\tilde{y}$.
\item[(o)] To obtain a credible interval for $l_T$, repeat (b) through (n)
$M_l$ times.
\end{enumerate}

\item\textit{Step} II: obtain the joint posterior predictive
distribution of wind characteristics $(v,s)$:
\begin{enumerate}[(a)]
\item[(a)] Find the MLEs of $\bolds{\nu}$ for all candidate
distributions listed in Section~\ref{subsecenv}.
\item[(b)] Use the $\mathrm{SIC}$ to select the ``best'' distribution model for
the average wind speed $v$. The chosen distribution is used in the
subsequent steps to draw posterior samples.
\item[(c)] Draw a posterior sample of $\bolds{\nu}$ from the
approximated multivariate normal distribution at the MLEs and the
inverse of the negative of the Hessian matrix.
\item[(d)] Draw $N_w$ samples of $\tilde{v}$ using the distribution chosen
in (b) with the parameter sampled in (c).
\item[(e)] Implement the RJS algorithm again, namely, (a) through (k) in
Step I, to get one posterior sample of $\bolds{\Psi}_{\eta
}=(\bolds{\beta}_\eta,\bolds{\phi}_\eta)$ and $\bolds{\Psi}_{\delta}=(\bolds{\theta}_\delta,\bolds{\phi}_\delta)$.
\item[(f)] Take the posterior sample of $\bolds{\Psi}_{\eta}$ and
$\bolds{\Psi}_{\delta}$, obtained in (e), and calculate a sample
of $\eta$ and $\delta$ using (\ref{eqns}) for each sample of $v$. This
generates $N_w$ samples of $\eta$ and $\delta$.
\item[(g)] Draw a sample for the standard deviation of wind speed $\tilde
{s}$ from each truncated normal distribution with $\eta_i$, $\delta_i$,
$i=1,\ldots, N_w$. Using the $N_w$ samples of $\eta$ and $\delta$
obtained in (f), we obtain $N_w$ samples of $\tilde{s}$.
\item[(h)] To get $M_w \times N_w$ samples of $\tilde{v}$ and $\tilde{s}$,
repeat (c) through (g) $M_w$ times.
\end{enumerate}
\end{enumerate}
In our implementation, we use $M_w=1000$, $M_l=10\mbox{,}000$, $N_{w}=100$
and \mbox{$N_l=100$}.

\section{Confidence intervals for the binning method}\label{secPROC}
To calculate the confidence intervals for the binning method, we follow
a procedure similar to the one used for calculating the credible
intervals in the spline method. The difference is mainly that in the
binning method, the parameters used in the GEV distribution, namely,
$\mu$ and $\sigma$ (recall that $\xi$ is fixed as a constant across all
the bins), are sampled using only the data in a specific bin. For those
bins which do not have data, its $\mu$ and $\sigma$ are a weighted
average of all nonempty bins with the weight related to the inverse
squared distance between bins, following the approach used by \citet
{AM2008}. Once a sample of $\mu$ and $\sigma$ is obtained for a
specific bin, the resulting local GEV is used to sample $\tilde{y}$ in
that bin. Do this for all the bins, and $\tilde{y}$'s from all bins are
pooled together to estimation $l_T$.

Specially, we go through the following steps, where $\bolds{\Phi
}_c$ denotes the collection of the parameters associated with all local
GEV distributions used in all bins:
\begin{itemize}
\item Draw $M_w \times N_w $ samples from the joint posterior
predictive distribution $p [\tilde{v},\tilde{s}|\mathcal
{D}_{v},\mathcal{D}_{s} ]$ of wind characteristics $(\tilde
{v},\tilde{s})$; this step is the same as in the spline method;
\item Using the data in a bin, draw a sample of $\mu$ and $\sigma$ for
that specific bin from a multivariate normal distributions taking the
MLE as its mean and the inverse of the negative of the Hessian matrix
as its covariance matrix. Not all the bins have data. For those which
do not have data, its $\mu$ and $\sigma$ are a weighted average of all
nonempty bins with the weight related to the inverse squared distance
between bins, as we explained above. Collectively, $\bolds{\Phi
}_c$ contains all the $\mu$'s and $\sigma$'s from all the bins;
\item Decide which bins the wind characteristic samples $(\tilde
{v},\tilde{s})$'s fall into. Based on the specific bin in which a
sample of $(\tilde{v},\tilde{s})$ falls, the corresponding $\mu$ and
$\sigma$ in $\bolds{\Phi}_c$ is chosen; doing this yields the
short-term distribution $p [\tilde{y}|\tilde{v},\tilde
{s},\bolds{\Phi}_c ]$ for that specific bin;
\item Draw $N_l$ samples of $\tilde{y}$ from $p [\tilde{y}|\tilde
{v},\tilde{s},\bolds{\Phi}_c ]$ for each of the total $M_w
\times N_w $ samples of $(\tilde{v},\tilde{s})$. This produces a total
of $M_w \times N_w \times N_l $ $\tilde{y}$ samples;
\item One can then compute the quantile value $l_T[\bolds{\Phi
}_c]$ corresponding to $P_T$;
\item Repeat the above procedure $M_l$ times to get the median and
confidence intervals of $l_T$.
\end{itemize}
Our implementation here uses the same $M_w, M_l, N_{w}$ and $N_l$ as
those used in the spline method's implementation.
\end{appendix}

\section*{Acknowledgments} This analysis has benefited from
measurements downloaded from the internet database: ``Database of Wind
Characteristics'' located at DTU, Denmark. Internet: \url
{http://www.winddata.com/}. Wind field time series from the following
sites have been applied: Roskilde, Denmark; Alborg, Denmark; and
Tehachapi Pass, California, USA. The authors would also like to
acknowledge the generous support from their sponsors.
%

%



\printaddresses

\end{document}